# Fabry-Perot Measurements of the Dynamics of Globular Cluster Cores: M15 (NGC 7078)

K. Gebhardt[1], C. Pryor[1], and T.B. Williams[1]

Department of Physics and Astronomy, Rutgers, The State University, Box 0849

Piscataway, NJ 08855-0849

gebhardt@physics.rutgers.edu, pryor@physics.rutgers.edu,williams@physics.rutgers.edu

and

James E. Hesser

Dominion Astrophysical Observatory, Herzberg Institute of Astrophysics, National

Research Council Canada

5071 W. Saanich Road, R.R.5, Victoria, B.C., V8X 4M6, Canada

hesser@dao.nrc.ca

Received ___________________;    accepted ___________________

astro-ph/9402064   24 Feb 94

## ABSTRACT

We report the first use of the Rutgers Imaging Fabry-Perot Spectrophotometer to study the dynamics of the cores of globular clusters. We have obtained velocities for cluster stars by tuning the Fabry-Perot to take a series of narrow-band (0.8 Å FWHM at 5890 Å) images at different wavelengths across one of the Na D (5890 Å) absorption lines. Measuring the flux in every frame yields a short portion of the spectrum for each star simultaneously. This proves to be a very efficient method for obtaining accurate stellar velocities; in crowded regions we are able to measure hundreds of velocities in 3-4 hours of observing time.

We have measured velocities with uncertainties of less than 5 km s$^{-1}$ for 216 stars within 1.5' of the center of the globular cluster M15 (NGC 7078). The velocity dispersion profile shows a sharp rise from 7 km s$^{-1}$ to 12 km s$^{-1}$ at 0.6' (1.8 parsecs), and then appears to flatten into our innermost point at 0.1'. A rotation amplitude of $1.4 \pm 0.8$ km s$^{-1}$ is detected. The rotation has been measured at a radius of 0.6' using stellar velocities and at 0.2' using the integrated light profile. The amplitude and position angle are the same at both radii, indicating a constant rotation profile in this region. Combining our two epochs of Fabry-Perot observation with published measurements, we have repeat velocity measurements for 67 stars. We calculate a binary fraction of about 7% for binary periods between 0.2 and 20 years and mass ratios larger than 0.22, which is in agreement with measurements for other globular clusters.

*Subject headings:* Globular Clusters, Stellar Systems (Kinematics, Dynamics), M15, NGC7078

## 1. Introduction

Measurements of the velocity dispersion profile, fraction of binary stars (both primordial and present-day), rotation, and velocity distribution moments are some of the more powerful tools for discriminating between different models for globular clusters. One of the interesting and important aspects of current studies of globular cluster evolution concerns the phenomena of dynamical core collapse (Hénon 1961, 1971; Spitzer & Hart 1971). We can hope to obtain an understanding of core collapse through observations of different clusters in various phases of this process. Detectable consequences of core collapse are limited to the central regions of the cluster, where the crowding of the stellar images becomes severe. In such crowded regions, users of single-slit, multi-slit and multi-fiber spectrographs face the problem of determining whether the light being measured originates from a single star or whether there is contamination from neighboring stars. In addition, fiber spacing limitations lead to observing inefficiencies in crowded regions.

An obvious alternative is to determine the velocity dispersion from the broadening of the integrated-light spectrum. But this approach is hampered by sampling errors since the light coming from the inner part of these clusters can be dominated by a few bright giants (Zaggia et al. 1992, Dubath 1993). This is not as significant a problem if measurements are made in the ultraviolet (King 1966), however most globular cluster stars emit little of their light at these wavelengths.

Peterson, Seitzer & Cudworth (1989, hereafter PSC) reported a rise in the velocity dispersion in the central region of M15 from 10 km s$^{-1}$ at 0.5$'$ to 25 km s$^{-1}$ in the inner 1$''$. Their result was obtained from velocities for 120 individual stars and, within the central 1$''$, from the broadening of the integrated-light spectrum. The apparent steep increase in the central velocity dispersion is suggestive of a central black hole and has consequently received considerable attention. Measurements of the accelerations of pulsars in M15 are more consistent with a value of the central dispersion of about 15 km s$^{-1}$ (Phinney 1993). Dubath, Mayor & Meylan (1993) has suggested that the PSC result may have arisen from sampling errors. The value of additional observational constraints is evident.

We present here 216 velocities for stars within 1.5′ of the center of M15 obtained using a new technique which can alleviate some of the problems due to crowding and sampling. The technique is discussed in section 2; Section 3 describes the observations and outlines the reduction procedure; and Section 4 gives the results, which are discussed in Section 5.

## 2. Technique

We have used the Rutgers Imaging Fabry-Perot Spectrophotometer on the CTIO 4-meter telescope to obtain images of globular clusters in a 0.8 Å FWHM ($\sigma = 40$ km s$^{-1}$ at 5890 Å) bandpass. These data can be analysed in two ways: to produce individual stellar velocities or a pixel-by-pixel velocity map. In the first approach, measuring the flux for every resolvable star in a series of images yields a short portion of the spectrum for each star in the vicinity of a strong absorption line. The images are spaced about 0.25 Å apart (one third of the FWHM) to provide good coverage of the absorption line profile. We use the point-spread function fitting program DAOPHOT II (Stetson 1987) to measure the fluxes.

The imaging Fabry-Perot can also yield completely sampled two-dimensional velocity information. A spectrum can be measured at each pixel, providing a velocity map of the cluster. Velocities at the positions of individual stars obtained in this way will be similar to those measured using DAOPHOT, but DAOPHOT does a much better job because it uses all of the light from a star, weighting each pixel by its signal-to-noise (S/N), and because it more accurately apportions the light between close neighbors in crowded regions. The advantage of a velocity map, however, is that it contains information on the integrated cluster light for those pixels which are not dominated by bright stars. In the crowded regions where sampling errors can be important, being able to avoid contamination from bright stars is necessary. Sampling errors are known to be significant for integrated-light studies of globular clusters (see, e.g., King 1966, Zaggia et al. 1992, Dubath 1993).

With FP data we can greatly reduce sampling errors by using DAOPHOT to subtract all identified stars from the image, thereby producing an image containing mostly diffuse cluster light from many fainter stars. The velocity map resulting from analysis of the diffuse cluster light will provide a better measure of the rotation profile than that obtained by using individual stellar velocities, since the integrated light has a smaller sampling uncertainty in the estimated mean velocity. We can also estimate of the velocity dispersion at each location in the cluster by measuring the broadening of the absorption line in the integrated light. The velocity dispersion estimate is particularly important within a few arcseconds of the cluster center, where dispersions calculated from individual stars are uncertain due to the small number of stars with measured velocities. Our velocity mapping procedure is highly dependent on the quality of the stellar subtraction process which, in turn, is dependent on the seeing. To date, regrettably, we have not obtained the seeing necessary to study the central velocity dispersion via the diffuse light, but, in principle, it can be a very powerful tool in studying the dynamics of the central few arcseconds of cusp clusters.

The typical total observing time per cluster is $3 - 4$ hours (using 15 10-minute exposures), reaching V = 17 in the uncrowded regions. The more luminous cusp clusters contain on the order of a few hundred stars this bright in our field. The initial observations discussed in §4.1 demonstrate that such data yield velocity uncertainties for individual stars that are about $1.0 - 5.0$ km s$^{-1}$, depending on the crowding and magnitude. The advantage of the FP technique is that we are able to get many velocities in crowded fields in a relatively short amount of observing time.

As a result of the optics of an imaging FP, light transmitted at the edge of the field in a given image will be bluer than light on the optical axis. The wavelength varies quadratically with radius, R, and for the Rutgers FP the wavelength gradient is 6 Å (at 5890 Å) across the 1.5′ radius of the field of view. Therefore, complete coverage of the absorption line at all radii would require approximately 32 images with our wavelength spacing. Since the wavelength is proportional to $R^2$, each additional exposure with the same wavelength spacing extends the coverage by equal-area annuli. Thus, the increase in the number of stars with adequate line coverage is simply given by the stellar number surface density, which goes as about 1/R in cusps, leading to a decreasing return for the investment of observing time. At some radius, the simultaneous wavelength coverage of multi-object spectra becomes more efficient. With a different optical design, the wavelength gradient could be lessened somewhat, and adequate line coverage could be obtained with less additional observing time. For the Rutgers FP, we get good coverage out to about 1′ from the cluster center with 15 exposures.

One of the serious problems with absorption-line FP work is establishing the frame-to-frame normalizations to account for atmospheric and instrumental transmission variations, since each sample of the profile is observed at a different time during the night or, possibly, even on different nights. We desire accuracies for the frame normalizations of better than 5%, which yield 2 km s$^{-1}$ velocity accuracies as estimated through Monte Carlo simulations. Techniques to handle non-photometric conditions will be discussed in §3.2. A desirable solution to the normalization problem in non-photometric conditions is to use a monitor channel for the transmission variations during the observations, but this is not an option for the current setup of the Rutgers FP.

## 3. Observations and Reduction

Below we present the specifics of the M15 observations and use the M15 data as an example of a typical FP reduction process.

### 3.1. Details of the M15 Observations

We used the CTIO 4-m telescope on July 9, 1991 and July 6-9, 1992 to obtain adequate line coverage for five clusters (M15, 47 Tuc, NGC 6397, NGC 6752, and M30). We report the results for M15 here. During the 1991 run we took 13 images of M15 with exposure times of 600 seconds each. The conditions were poor: non-photometric with transparency variations of up to 40% and seeing that varied from $2.0'' - 2.4''$ FWHM. During the 1992 run we obtained 15 frames of the same exposure time, taken in conditions that were not that much better: transparency variations of up to 98% and seeing that varied from $1.5'' - 2.0''$. Due to the poor seeing during the 1991 run, the images were binned on-chip to 2x2 ($0.8''$ pixels), but for 1992 they were unbinned ($0.4''$ pixels).

The 1991 images were centered in wavelength on the red Na D line (5896 Å) and the 1992 images on the blue Na D line (5890 Å). We sampled in wavelength equally on both sides of the absorption line for an object at the center of the cluster. For the 1991 data, the wavelength gradient in the FP gave us the benefit of scanning across the blue Na D line for stars on the edge of the frame since the center-to-edge gradient in the FP is about 6 Å, which is the same as the difference in wavelength between the two Na D lines. We were then able to measure velocities for stars both in the center of the cluster and at the outer edge of the frame. This was not the situation for the 1992 data, but, since the seeing was better then, we were able to measure more velocities.

The spectra show absorption from the interstellar medium (ISM), but, due to the high negative velocity of M15 ($-107$ km s$^{-1}$ as measured by PSC), the stellar line and the ISM absorption line are not blended. The velocities and equivalent widths of the ISM lines were measured from 120 stars with high S/N and then held fixed at the average values when fitting the stellar line. There are two identifiable velocity systems of the ISM, and our velocities and equivalent widths for those systems agree with previous measurements (Cohen 1979, Songalia & York 1980, Kraft & Sneden 1993). Our data show obvious changes in the velocity and equivalent width of the ISM lines which are not pursued in this analysis, but could be used to map the small-scale variation of the ISM properties.

## 3.2.  Reduction of the M15 Data

Our reduction procedure includes measuring the etalon drifts, flatfielding, removing reflections (ghosts), determining the stellar flux, fitting the line profile, and determining frame-to-frame normalizations.

Both the wavelength zero-point of the FP and the center of the optical axis drift slowly with time. To monitor this, we take calibration rings every hour or whenever the telescope is moved. We use both an Argon and a Neon lamp for the calibrations. The drift is typically about 0.05 Å per hour for the zero-point, and about 0.25 pixels per hour for the optical center.

Since the FP passes many interference orders, we must use a pre-filter to select the order that we want. The free spectral range of the etalon is about 18 Å, so the pre-filter must be narrow (about 15 Å) and have sharp edges in order to minimize contributions from other orders. The response of this pre-filter is not flat over the portion of the spectrum scanned in these observations and this response must be carefully removed. This effect is particularly significant for wavelengths close to the edge of the pre-filter passband. Thus we take a flat field exposure at nearly every wavelength setting (to within 0.1 Å) used during the observations to ensure that we can properly flatten each object frame. The flats were taken with a projector lamp, and the illumination correction for the projector was determined by comparison of projector and dome flats taken at a single wavelength setting.

Some light reflected from the CCD also reflects from the FP to form nearly in-focus ghost images with mirror symmetry about the optical axis. An anti-reflection coating on the CCD reduces the amplitude of the ghosts to about $2-3\%$, but we still remove the ghost images with a de-ghosting routine. In order to de-ghost an image we must know the relative reflectivity at each pixel. We determine this reflectivity map by assuming that the light that is absorbed by the CCD is also detected. Since light is either absorbed by or reflected off of the CCD, the reflectivity map is the complement of the flat-field image (reflectivity = 1 – absorbed). We apply a Gaussian spreading (to compensate for the defocussing upon reflection) to the reflected light and an overall scale factor, where these parameters are determined empirically. We can then subtract the ghost images and determine the total light intensity which originally illuminated each pixel.

We determine flux values for each star on each frame using DAOPHOT II (Stetson, 1987). We typically use about 50 bright, isolated stars to determine the PSF on every frame. The same PSF stars are used in each frame and the PSF is allowed to vary quadratically across the frame.

Due to severe crowding in the central regions of our images, it is important to have good positions for the stars in order for the DAOPHOT deconvolution to work more precisely. Fortunately, Yanny et al. (1993) have measured the positions of stars in the M15 core with Hubble Space Telescope (HST hereafter). Since these data only extend to about $0.6'$ from the cluster center, we have used DAOPHOT to determine positions for stars outside of that region. The HST positions for the central $0.6'$ and the DAOPHOT ones for the remainder of the field provide our initial star list. Using our seven best-seeing frames, we allowed DAOPHOT to determine the position for each star in each frame from the input initial star list, and then averaged the positions from the seven frames. These final star positions are then kept fixed when determining the final magnitudes for each star, taking into account any spatial offsets between frames.

The instrumental profile of the FP is well fit by a Voigt function. This convolution of a Gaussian and Lorentzian (Cauchy) distribution is given by

$$V(x, y) = \frac{1}{\sqrt{2\pi}\,\sigma_g} \frac{y}{\pi} \int_{-\infty}^{\infty} \frac{e^{-s^2}\,ds}{y^2 + (x - s)^2} \quad .$$

Here

$$x = \frac{x' - x_o}{\sqrt{2}\,\sigma_g}, \quad \text{and} \quad y = \frac{\sigma_l}{\sqrt{2}\,\sigma_g},$$

where $\sigma_g$ is the scale (e.g., standard deviation) for the Gaussian, $\sigma_l$ is the scale for the Lorentzian, $x_o$ is the central location (e.g., mean) and, $x'$ is the wavelength of interest. The evaluations of the Voigt function and the derivatives are done through a polynomial expansion (Humlíček 1979, Varghese 1992). This method has been compared to the actual evaluation of the integral using Fourier transforms (using the routine GALFIT; see Ouaynq & Varghese 1989), and the relative difference is less than $10^{-6}$, which is well within our desired tolerance. We determine the instrumental parameters of the Voigt function profile from approximately 30 spectral lamp calibration frames. When fitting a stellar spectrum, the intrinsic width of the stellar line is added in quadrature to the Gaussian scale of the instrumental response. We then have a four parameter fit to each spectrum: the continuum level, the equivalent width, the central value, and the scale for the stellar line.

We fit the profile using least-squares, employing the Levenberg-Marquardt method from Numerical Recipes (Press et al. 1986). An initial fit uses as the uncertainties the $1\sigma$ errors given by DAOPHOT. However, to ensure that the fitted profile is not biased by other weak lines in the outer parts of the line profile, and also to reduce the weight of the contribution from the ISM lines, we then re-weight the flux values according to their position from the calculated central value added in quadrature with the DAOPHOT errors, and re-fit. All points within 0.7 Å ($\sim$1 FWHM) of the fitted central value are given the same weight, while for points outside of that interval the weights decrease quadratically with distance from the central value. This procedure is repeated and converges in two or three iterations.

Since our observing conditions were not photometric, we use stars on the edge of the field for initial determination of the frame-to-frame normalizations. At the field edge, the wavelength is 6 Å bluer than that at the optical center (at 5890 Å), so the spectra of stars near the edge are in the continuum and we simply normalize the frames to a flat spectrum. But if the absorption line is very broad or if weaker lines are present blueward of Na D, the continuum may, in fact, not be flat, in which case our procedure will introduce systematic errors larger than our desired accuracy (5%). Fortunately, we can use additional information to refine our normalizations: stars having independently-measured velocities. For each such star we fit a line profile while holding the velocity fixed at its independently-measured value and use the deviation of the individual points from the fitted profile as estimates of the normalizations for each frame. For M15, we use 45 stars with measured velocities from PSC and adopt as the normalization for each frame the central location of each set of 45 deviations. The fits are then repeated and converge in a few iterations. The above procedure will not create a line where none is present because these 45 stars have a range of radii in the cluster, which means that the spectral line occurs in different frames for different stars. The different velocities of the stars also shift the line to different frames, but to a lesser extent.

Throughout the reduction procedure and analysis, we estimate the central location (mean) and scale (standard deviation) of a distribution by using the biweight, which is a robust estimate, insensitive to outliers; it is needed to ignore possible large deviations caused by stars with cosmic rays superimposed on their image or with variable velocities. For a description of the biweight estimator see Beers, Flynn & Gebhardt (1990). The biweight estimate of the scale of the 45 deviations is our measure of the the accuracy of our normalization procedure, which for M15 is better than 2%.

We find that even after the above procedure there are small zero-point offsets between our velocities and previous measurements. Such an offset may be a result of a small systematic error in the frame normalizations, or an error in the calibration of the wavelength zero-point of the FP. In any case, this will not significantly affect our relative velocities. Any measured zero-point offset, which is typically less than 1 km s$^{-1}$, will be added to our velocities.

## 4. Results

### 4.1. The Velocities and Their Uncertainties

We have measured velocities for 216 stars within the central 1.5′ of M15. The results are given in Table 1. The center of cluster was taken to be $21^h27^m33^s.3, +11°56'49''$ (1950), as measured by HST (Lauer et al. 1991). The columns are the ID, either taken from Küstner (1921) or Aurière & Cordoni (1981) if the star has been previously studied, or an FP number if it has not (col. 1), the x and y offsets in arcseconds from the cluster center (cols. 2 and 3), the FP V magnitude (4), the velocity and its uncertainty (5), the source of the velocity measurement (6), the probability of the $\chi^2$ of the multiple velocity measurements (see §4.2) (7), and other ID's and references to comments for individual stars (8). We have included all of the measurements from PSC in Table 1 for ease of comparison. The other ID's for stars are as follows: V is a variable star number (Sawyer-Hogg 1973); S is the ID given by Sandage (1970); and a roman numeral is the ID from Arp (1955). When there are multiple velocity measurements of the same star, the first line for that star lists the average velocity calculated from all of the measurements weighted by their uncertainties and the following lines are the individual measurements. The FP V magnitude is determined from the fitted continuum level, which was calibrated with the HST V magnitudes. The rms scatter between the HST magnitudes and the FP magnitudes is 0.16 magnitudes. Our magnitudes should only be used as rough estimates of the V magnitude since we only have 4 Å of spectrum. Table 1 is presented in its complete form in ApJ/AJ CD-ROM Series, volume 2, 1994. The first page of this table is presented here for guidance regarding its form and content.

In Fig. 1 we plot line profiles for three stars in M15, taken from the 1992 data. Each point is the flux (in electrons) measured for that star in a frame. The error estimates for each point are those calculated by DAOPHOT. Each stellar spectrum does not have the same wavelength range because there is a gradient in wavelength from the center to the edge in the FP and the stars are at different distances from the FP center. The solid line is the fitted Voigt profile. The FP magnitudes of the stars from top to bottom are 13.2, 14.0, and 16.0. The ISM absorption line at 5889.7 Å is clearly visible, and is actually stronger than the stellar line, but, due to the velocity difference, our derived velocities for M15 stars are unaffected (in part because the fitting is done by strongly weighting the points in a $\pm 0.7$ Å region around the line central location).

Our velocity uncertainties are determined by Monte Carlo simulations. We use the DAOPHOT $1\sigma$ error estimates and assume that the flux values are normally distributed. In addition, the calculated frame normalizations are assumed to have a normal distribution with the standard deviation given by their estimated uncertainties. For M15, the normalization uncertainties were taken to be 2% . We generate 1000 realizations of the spectrum for every star and fit a profile to each. We adopt the biweight scale estimate of the resulting velocity distribution as the velocity uncertainty.

Since the velocity dispersion of M15 is about 10 km s$^{-1}$, we have chosen to report in Table 1 only velocities which have uncertainties less than 5 km s$^{-1}$. This conservative cut ensures that our measured velocity dispersions are not strongly dependent on our estimated uncertainties. We measure velocities for approximately 80 additional stars with uncertainties between 5 km s$^{-1}$ and 10 km s$^{-1}$. We also apply two other cuts to our sample. We use HST positions for stars in the M15 core to determine whether the measured light of a star in our sample has been significantly contaminated by neighboring stars. HST has resolved every star down to a limiting magnitude of about V=19; we use the HST positions and magnitudes to calculate, for every star for which we have a measured velocity, the fraction of light that other stars contribute to the total light inside one DAOPHOT fitting radius (1.6″). We then calculate this fraction in the same manner using the stars which have been identified in the FP frames, using the magnitudes from the fitted profile. If a star did not have a fitted profile, the estimated magnitude was an average from four frames which were known to be in continuum. This gives us the contamination fraction from HST and from the FP data. If the FP contamination fraction does not agree with the HST value to better than 20% for every contaminating star, we do not accept the measured velocity for the star being considered. This cut typically eliminates about 10% of the stars with velocities formally more accurate than 5 km s$^{-1}$. We stress that this is a problem which affects all types of observations in crowded regions, and is of particular importance to slit spectra, where two dimensional information is not available to properly assign contributions to the light of a particular "star". Our final cut consists of visually examining each profile to decide whether it is believable or not; this procedure eliminated less than 5% of the stars, and is most helpful for cases where the wavelength coverage of the line profile is inadequate.

Although Table 1 includes velocities for 289 stars, we cannot use all of these to study the cluster kinematics. As noted in §2, the gradient in wavelength in the FP field causes the wavelength coverage to be shifted blueward for stars at larger radii; thus at larger radii, stars with velocities more negative than the cluster mean are more likely to have complete profile coverage and so be included in our sample, while stars with positive velocities are more likely to have incomplete coverage and so be excluded. For example, for stars at the optical center (4″ from the cluster center), we are able to measure velocities between –60 km s$^{-1}$ and +90 km s$^{-1}$ from the cluster mean; for stars at 0.7′ from the optical center, we can measure between –120 km s$^{-1}$ and +30 km s$^{-1}$; and outside of 1′ we can only measure velocities more negative than –40 km s$^{-1}$. There have been several reports of stars with unexpectedly high velocities in globular clusters (Gunn & Griffin 1979, Meylan et al. 1991), which are important to quantify for understanding the cluster dynamics. Depending on the velocity and position of a star, this setup of the FP may not be able to measure velocities for some of these stars. But, in the central regions (inner 0.5') where high-velocity stars have been found in other clusters, we have adequate coverage to measure their velocities. In fact, for the two high-velocity stars measured by Meylan et al. 1991 in 47 Tuc, we were able to obtain the same velocity measurements (Gebhardt et al. 1994).

The bias in our sample from the wavelength gradient in the FP must be incorporated for a proper dynamical analysis and there are two possible approaches. We could use analysis techniques which consider truncated datasets with a known bias, or we could exclude any data that are outside of the radius where the bias begins to have a significant effect. Due to ease of initial computation and understanding, we choose the latter. Thus we exclude stars further from the optical center than the point where a velocity that is more than three times the dispersion more positive than the cluster mean would not be measured by the FP. Such stars are flagged with a '2' in the Note column of Table 1. Combining the remaining FP velocities with the previous measurements of PSC yields 245 stars which we use to compute the cluster dispersion and the rotation.

The 1991 data yielded velocities for 71 stars and the 1992 data yielded velocities for 191. Fig. 2 plots our 1992 velocities vs. those from 1991. There are only 46 stars in common because of different seeing and because different absorption lines were used (see §3.1). There is an offset of 1.9 $\pm 0.6$ km s$^{-1}$ and the scatter about the offset is 4.1 km s$^{-1}$. The dashed line represents the measured offset of the two samples. This offset is more than expected (we typically have offsets less than 1 km s$^{-1}$, see §3.2), and we therefore use previously measured velocities to determine our overall offset for the 1991 and the 1992 datasets. PSC have measured velocities for stars in the central region of M15 and from the comparison we find offsets of $-0.6 \pm 0.5$ km s$^{-1}$ from the 1991 velocities of 30 stars and $1.0 \pm 0.4$ km s$^{-1}$ from the 1992 velocities of 45 stars (where $V_{\text{off}} = V_{\text{PSC}} - V_{\text{FP}}$). Fig. 3 plots our measured velocities vs. those of PSC, for the 47 stars in common. The determined offsets have been added to our velocities. When a star had a measured velocity from both 1991 and 1992, we have used the average weighted by the variances. The scatter about zero of the differences between the PSC velocities and the FP combined velocities is 2 km s$^{-1}$. The distribution of the velocity differences divided by their uncertainties approximates a normal distribution with variance one for all comparisons, which gives us confidence that our error estimates are reliable. The star with the largest discrepancy in Fig. 2, about 20 km s$^{-1}$, is AC5, which is the photometric variable V86 (Sawyer-Hogg 1973). Other stars with discrepant velocities, which are candidates for binary stars, will be discussed in §4.2.

Fig. 4 plots the average radial velocities (the first entry for each star in Table 1) versus the distance of the star from the center of the cluster. Different symbols show stars measured by PSC alone, by us alone, or by both groups. The FP data increase the number of stars inside a radius of 1′ by 4.5 times and nicely complement the PSC velocities measured with a single-slit spectrograph at MMT. We were able to measure very few velocities in the central 0.1′ of M15 because of the crowding and our poor seeing. There are two stars in the central region, AC224 and AC104, for which PSC have reported a velocity and we do not. Although we have measured a profile for each of these, they were determined from the HST positions and magnitudes to have significant contamination from neighboring stars.

Ignoring this, PSC's velocity for AC224 is $-75.4$ km s$^{-1}$ while ours is $-83.4$ km s$^{-1}$. For AC104, the two velocities are $-97.6$ km s$^{-1}$ and $-106.7$ km s$^{-1}$, respectively. The large velocity differences for both stars is another indication that there are concerns about the reliability of these measurements. These stars are marked with a '3' in the Note column of Table 1 and are not used in the following analysis of the dispersion and the rotation.

## 4.2. Binaries

Including the two FP runs and the PSC data, there are 69 stars which have repeat velocity measurements. For each star with a repeat measurement we have calculated the weighted mean velocity from the individual measurements. We then calculate the $\chi^2$ of the scatter of the velocities about the mean and give in column 7 of Table 1 the probability of that $\chi^2$ value being exceeded, assuming that the velocity is constant. There are two stars with probabilities smaller than 0.001: AC5, which has two FP observations separated by 20 km s$^{-1}$, and K673, which has one PSC and two FP observations with a range of 14 km s$^{-1}$. A third star, AC161, has a $\chi^2$ probability of 0.008 and a velocity range of 11 km s$^{-1}$ with three observations. AC5 is the pulsating photometric variable V86 (Sawyer-Hogg 1973) and so is not a binary. The velocity ranges for K673 and AC161 are larger than that resulting from the velocity "jitter" of stars near the giant branch tip (Mayor et al. 1984, Lupton et al. 1987, Pryor et al. 1988). Thus we consider K673 and AC161 to be good binary star candidates and they are labelled with a '1' in the Note column of Table 1. Additional photometric and radial velocity observations of these stars are needed to confirm that they are binaries. We use the weighted mean velocities of the velocity variables in the dynamical analysis.

Though our sample of stars and the number of repeat velocities are small, we briefly compare our results with other recent binary surveys. After eliminating several stars with repeat measurements separated by a few days or less (and treating the PSC radial velocity standard K144 as having one velocity at each of the 4 observing epochs), we have a sample of 66 stars with 184 velocities. The average time baseline is 4.0 years and 25 of the stars have 3 or more velocities. Unlike other studies (e.g., Pryor et al. 1988), we do not find any tendency for the brighter stars to have less probable $\chi^2$'s. This could be due to the somewhat larger uncertainties of the FP velocities, however. We agree with other studies (Pryor et al. 1989, see the review in Hut et al. 1992) in finding that our binary candidates are not significantly centrally concentrated when compared to our complete sample.

We have simulated our observations for populations of binaries with different periods, mass ratios, and orbital eccentricities ($e$) (see Hut et al. 1992 for a more detailed discussion of the procedure). For binaries with periods between 0.2 and 20 years and mass ratios larger than 0.22, using a discovery criterion that the $\chi^2$ probability be less than 0.001 with our data would find 0.38 of the systems assuming circular orbits and 0.15 of the systems assuming an eccentricity distribution $f(e) = 2e$. These discovery efficiencies are similar to those of other studies. Our discovery fraction of $1/66 = 0.015$ then implies a binary frequency for the above range of periods and mass ratios of 0.04 or 0.10, depending on the unknown eccentricity distribution. The 95% confidence intervals are 0.02–0.21 and 0.06–0.53, respectively. These confidence intervals were calculated using the binomial distribution (see Hut et al. 1992). These values are consistent with the corresponding frequencies of 0.05 and 0.12 for six clusters (Hut et al. 1992), of 0.1 and 0.2 in NGC 3201 (Côté et al. 1993), and of 0.15 and 0.27 in NGC 362 (Fischer et al. 1993).

### 4.3. Velocity Dispersion

The velocity dispersion profile is plotted in Fig. 5. The small filled circles are the absolute value of each star's deviation from the cluster velocity plotted vs. the distance from the center. The solid line and the open circles with error estimates are the velocity dispersion estimated by two different techniques. The open circles are a maximum likelihood estimate of the dispersion (Pryor & Meylan 1993) in bins of 22 stars. This method removes the contribution of the measurement uncertainties. The solid line is a locally weighted scatterplot smoothing (LOWESS, see Cleveland & McGill 1984) fit to the velocity deviations squared. This technique estimates the velocity variance at each data point using the straight line fit by least-squares to the individual squared deviations in a kernel around that point, weighting the points by the inverse square of the distance in radius from that point. The square root of the fitted variance is an estimate of the velocity dispersion. The dashed lines are the boundaries of the 90% confidence interval estimated through Monte Carlo simulations. We construct synthetic data sets by randomly choosing a velocity from a Gaussian distribution with the standard deviation given by the dispersion profile at the radius of each observation and the uncertainty of each velocity measurement. We do a LOWESS fit to each of 1000 realizations and then estimate the 90% confidence limit from the resulting distributions. The two techniques give the same profile to within the uncertainties, which gives us confidence that the shape of the dispersion curve is robustly determined. Within $0.1'$ there are too few data points to be able to say anything about the dispersion. The LOWESS fit continues to the innermost data value, but it should not be trusted.

PSC presented evidence from integrated light measurements of the central cusp that the velocity dispersion rises sharply in the central $1''$ of M15 (from 10 km s$^{-1}$ at $30''$ to 25 km s$^{-1}$ at $1''$). Because of our $1.8''$ seeing we cannot determine the dispersion accurately within the central $1''$ and, hence, we cannot confirm this interesting result. Our calculated dispersion at $30''$ is 10 km s$^{-1}$ and the dispersion reaches its maximum, of only 12 km s$^{-1}$, at $20''$. At smaller radii the dispersion remains constant. Our data confirm the rise seen by PSC between $0.7'$ and $0.4'$, but give dispersion estimates about $1.7\sigma$ lower in the region between $0.1'$ and $0.4'$. Inclusion of the PSC velocities for AC104 and AC224 (which were excluded earlier due to contamination) changes the result insignificantly; with them included, the velocity dispersion of the innermost bin at $0.14'$ increases from 9.5 km s$^{-1}$ to 11.2 km s$^{-1}$, which is within the $1\sigma$ error estimates.

### 4.4. Rotation

Detecting a net rotation of a few km s$^{-1}$ in a cluster with an internal velocity dispersion of 10 km s$^{-1}$ is a challenging goal, as Fig. 6 demonstrates; fundamentally, it is a problem that can only be solved with many velocity measurements. The deviation of each velocity from the cluster mean velocity is plotted vs. the position angle on the sky (PA is measured from N through E) for the 245 stellar velocities. The solid curve is the sine curve that best fits the data. The fitted position angle of the rotation axis is $295° \pm 33°$ and the amplitude is $1.4 \pm 0.8$ km s$^{-1}$. Monte Carlo simulations indicate that there is a 5% chance of measuring a 1.4 km s$^{-1}$ rotation amplitude assuming that no rotation is present. Thus we apparently do detect significant rotation. Using a smaller sample of 120 measurements, PSC did not find significant rotation in M15.

We can do better by using the line profile of the integrated cluster light to estimate the rotation. We fit a line profile to every pixel in the data frames, which produces a velocity map of the cluster. We then bin these velocities in azimuth and estimate the central location within each bin. Fig. 7 plots these velocity estimates vs. position angle. The error estimates are the $1\sigma$ uncertainties estimated from the scatter of points in each bin. The solid line is the best-fit sine curve to the integrated light data and the dotted line is the sine curve from Fig. 6, which was determined from the individual stellar velocities. The fitted position angle of the rotation axis is $310° \pm 11°$ and the amplitude is $1.7 \pm 0.3$ km s$^{-1}$. The two sets of data are not completely independent since some pixels contain stars with measured velocities, but this only affects 10% of the pixels. The measurement from the integrated light is restricted to pixels within $15''$ of the cluster center, since the cluster light becomes too weak for reliable velocity measurement beyond $15''$. However, the majority of stars with measured velocities are at larger radii and so the fit to them is more representative of the rotation at their average radius of $36''$. From Fig. 7, it is apparent that neither the PA nor the amplitude of the rotation changes significantly with radius.

Since the center of the FP and the cluster center are not at the same location (they are different by $4''$), it would be possible to create a rotation signature if the calibration parameter for the wavelength gradient is not correct. For the velocity difference of 4 km s$^{-1}$ measured by the integrated light, the calibration parameter would have to be in error by 30%, while we have determined it with an estimated precision of 1%. In addition, the position angle of the rotation is not aligned with the position angle of the line connecting the FP and cluster centers. Thus the rotation is an actual feature and not an artifact of the reduction and calibration.

## 5. Discussion

Fig. 5 shows that the velocity dispersion increases from 7 km s$^{-1}$ to 12 km s$^{-1}$ at a radius of about $0.6'$. This is consistent with the rise in the dispersion measured by PSC at the same radius. However, from $0.4'$ to $0.1'$, our data suggest a constant velocity dispersion of about 11 km s$^{-1}$, whereas PSC measured an increasing dispersion in this region. The dispersions measured by PSC at $0.1'$ and $0.2'$ are above our estimated 90% confidence interval at those radii, which is most likely a result of our increasing the sample size in this region by a factor of two. Within $0.1'$ we have inadequate data to measure the velocity dispersion and we therefore cannot directly compare our central dispersion estimate of 11 km s$^{-1}$ with the 25 km s$^{-1}$ value reported by PSC.

PSC had suggested that the steep rise in the dispersion in the central $1''$ was due to a black hole in the center of the cluster. However, Grabhorn et al. (1992), using Fokker-Planck modelling, were able to fit models with a high central velocity dispersion without the need for a central black hole, although they prefer a model with a central dispersion of 14 km s$^{-1}$. This latter value agrees with the measurement of Dubath, Meylan & Mayor (1993) from the integrated light in the central $6'' \times 6''$ and with the constraints from the pulsar accelerations (Phinney 1993). Our measured value at $6''$, 11 km s$^{-1}$, is also consistent with the preferred model value, although slightly lower. Since we do not have enough velocities to reliably estimate the dispersion in the inner $6''$, our data neither rule out the possibility of a central black hole, nor do they provide any evidence for one.

Our data suggest that there is a shoulder in the velocity dispersion profile at a radius of about $0.4'$. This will be a significant obstacle for the current models (both multi-mass King-Michie and Fokker-Planck) to fit. An alternative approach to modelling is to use the velocity data and the surface brightness data to make non-parametric estimates of the mass density profile of the cluster (Merritt 1993, Merritt & Saha 1993). These techniques are only possible with large datasets (500-1000 stars), which we are now beginning to approach. We have used non-parametric fitting to determine the mass profile of M15 by direct inversion of the Jean's equation assuming isotropy. At the location of the shoulder in the velocity dispersion profile, the mass density profile also exhibits a shoulder. This shoulder is suggestive of mass segregation in the inner regions. This will be discussed fully in a future paper (Gebhardt & Fischer 1994). Mass segregation has been seen with star counts (Richer & Fahlman 1989), but we are now in a position to directly and strongly constrain the mass profile, the amount of mass segregation, and the stellar remnant population from the velocity and surface brightness data.

The rotation signature evident in Fig. 7, seen in both the individual stellar velocities and in the integrated light, indicates that the rotation amplitude varies little over the range of radii sampled; the average radius for the stars is $36''$, while the average radius for the pixels contributing to the integrated light measurement is $8''$. This lack of decrease in the rotation amplitude is not consistent with the solid body rotation which one might naively expect in the dense central region of M15 because of the short two-body relaxation time there (the half-mass relaxation time is 2.7 Gyr, see Grabhorn et al. 1992). Models which have been used for the rotation of globular clusters assumed a decreasing rotation amplitude with decreasing radius in the central region (Davoust 1986; Meylan & Mayor 1986; Fischer, Welch & Mateo 1992). Akiyama & Sugimoto (1989) used N-body calculations to model the evolution of rotating star clusters and found that, assuming an initial condition of constant angular frequency, the rotational velocity decreased with decreasing radius at all times. Although the rotation amplitude of 2 km s$^{-1}$ may not be dynamically significant compared to the dispersion of 11 km s$^{-1}$ for the central part of the cluster, the constancy of the rotation amplitude with radius suggests interesting dynamics in the central region. Peterson (1993) measures a possible rotation amplitude of 7.5 km s$^{-1}$ in a $0.5''$ x $5''$ slit near the cluster center using the H$\alpha$ profile of the integrated light, which would have important dynamical consequences.

We identified three stars with significantly variable velocities among the 67 stars in our sample with multiple observations separated by more than a few days. One is a photometric variable and the remaining two are good candidates for binary stars, though more observations are needed to confirm their nature. The candidates are not significantly concentrated towards the cluster center when compared to our total sample (though we emphasize that our sample is small). The frequency of binaries in M15 agrees within its order of magnitude uncertainties with those of the other globular clusters where it has been measured.

We are grateful for discussions with many people during the observations, reduction and analysis: Peter Stetson for many conversations about, and programs for, the photometry; Philip Varghese and Xiang Ouayng for the Voigt function fitting routines and suggestions on how to use them; Bob Kraft and Chris Sneden for sending unpublished spectra of M15 stars; M. Aurière for providing his list of M15 stars; Raja Guhathakurta and Brian Yanny for providing HST data before publication; Mauricio Fernandez for his excellent help at the telescope; and the rest of the CTIO staff for their outstanding assistance throughout. The Rutgers FP was developed with support from Rutgers University and the National Science Foundation, under grant AST-83-19344. CP and TW acknowledge support for their globular cluster research through grant AST-90-20685 from the National Science Foundation. KG wishes to acknowledge support from the U.S. Department of Education. Data available upon request from KG.

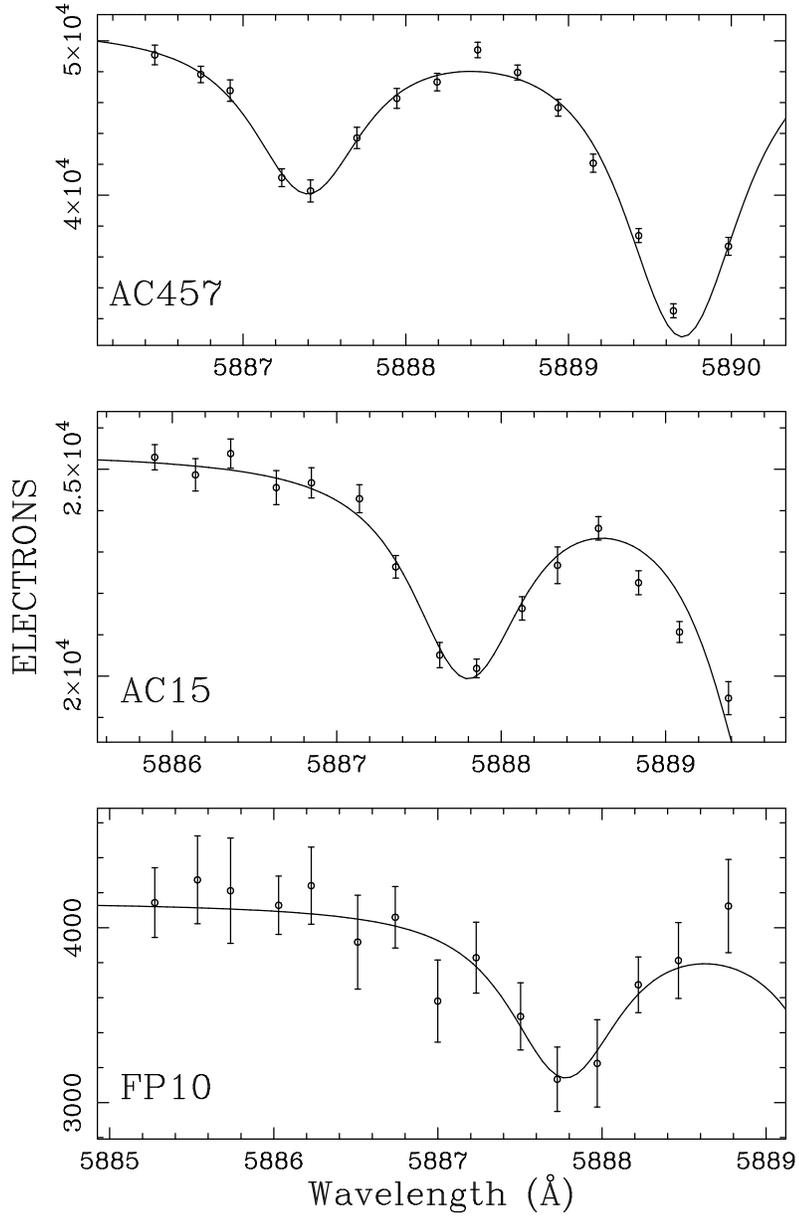

Fig. 1.— Profiles for three stars of magnitudes $V_{FP} = 13.2$, 14.0, and 16.0, from top to bottom. The line at about 5887.7 Å is the stellar absorption line and the line at 5889.7 Å is the ISM absorption line.

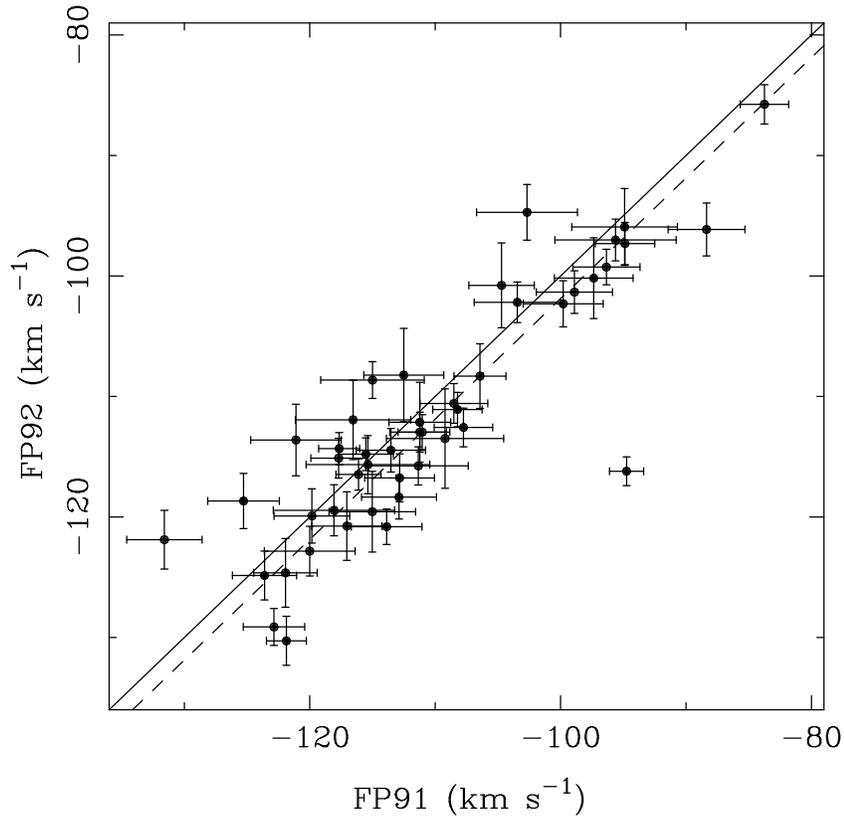

Fig. 2.— Comparison of the 1991 and 1992 FP velocities for the stars in common. The solid line is equal 1991 and 1992 velocities and the dashed line is the best-fit offset determined from the average differences ($V_{FP91} - V_{FP92} = 1.9 \pm 0.6$ km s$^{-1}$).

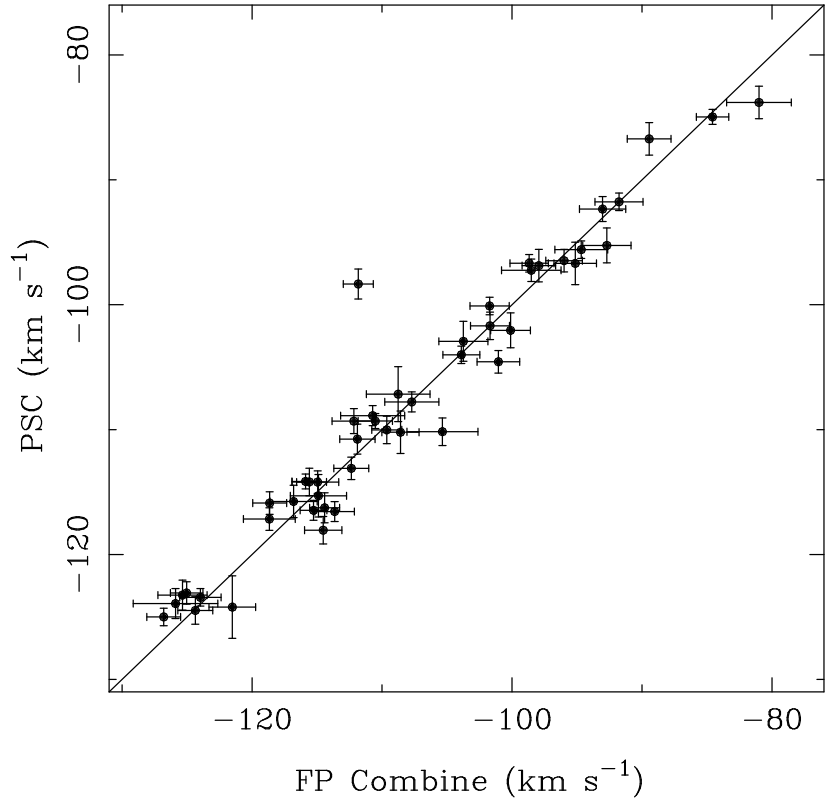

Fig. 3.— Comparison of PSC and FP velocities for the stars in common. If a star had a FP measurement from both 1991 and 1992, the average velocity weighted by the uncertainty was used. Velocity offsets were added to both the 1991 and 1992 data to make them agree with the PSC cluster velocity.

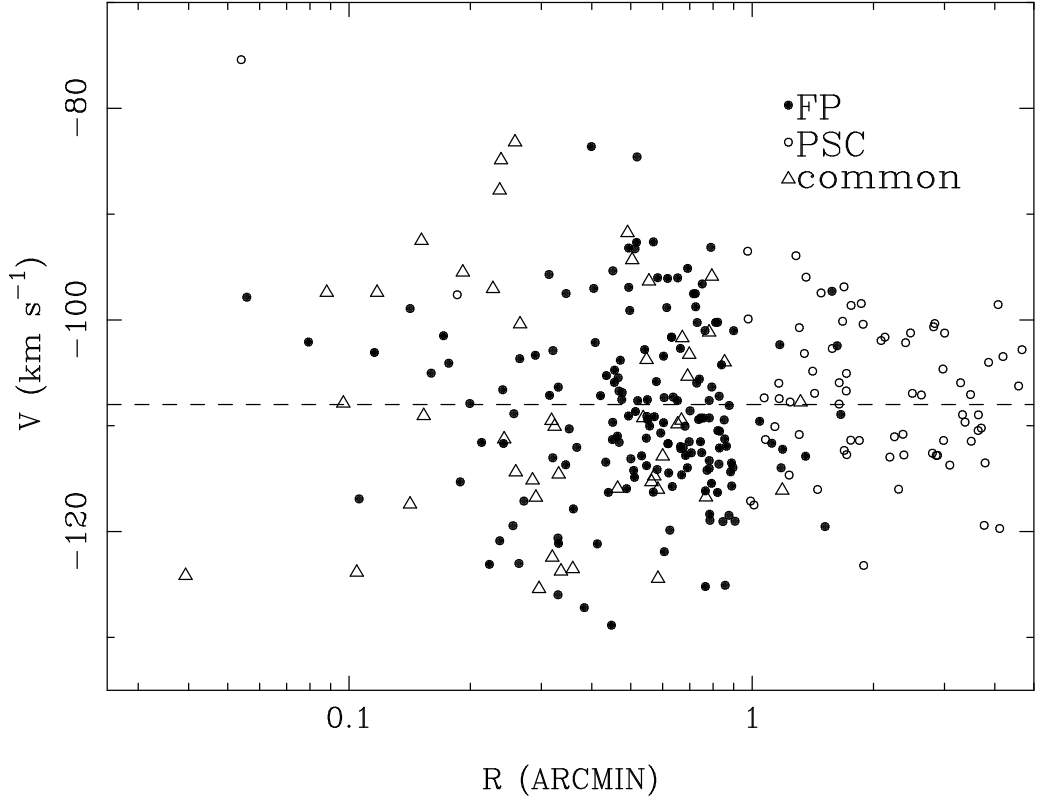

Fig. 4.— Velocity vs. radius from the cluster center for the 245 stars which have been used to study the kinematics of M15. The dashed line is the measured cluster mean velocity ($-107.9 \pm 0.5$ km s$^{-1}$).

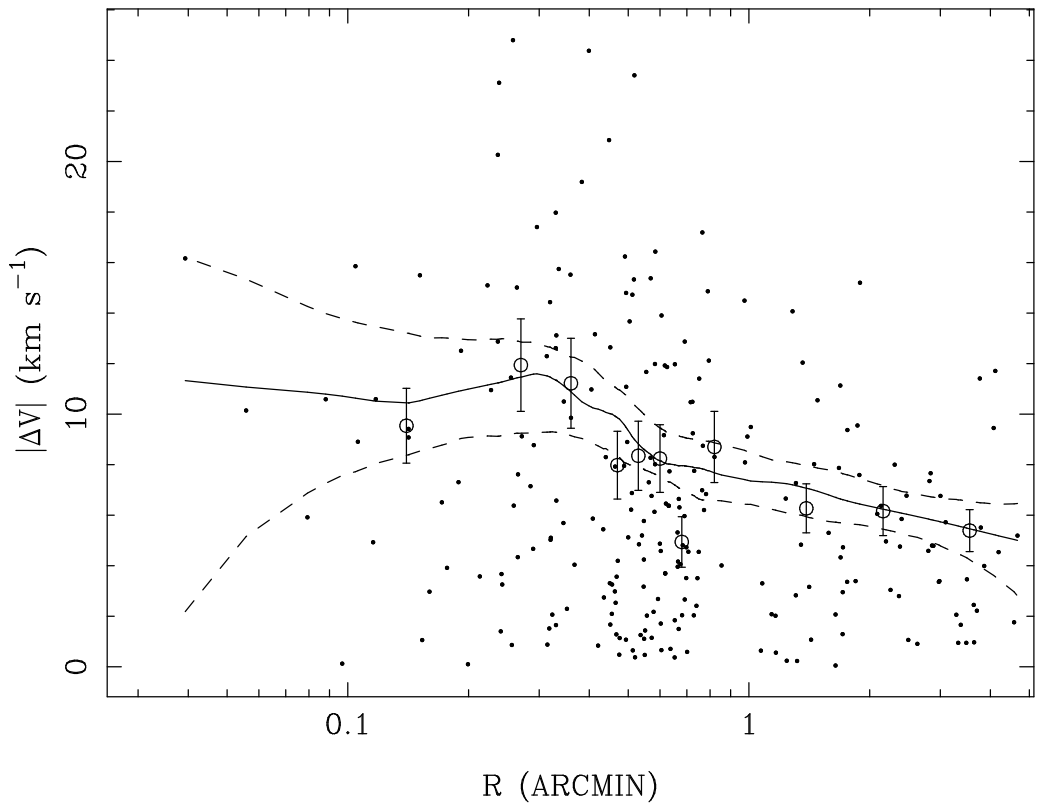

Fig. 5.— Velocity dispersion vs. radius for stars in M15. The filled circles are the absolute deviations of the individual measurements from the cluster velocity plotted vs. radius from the cluster center. The open circles are the velocity dispersion estimates, with uncertainties, in bins of 22 stars. The solid line is the LOWESS estimate of the dispersion and the dashed lines are its 90% confidence interval.

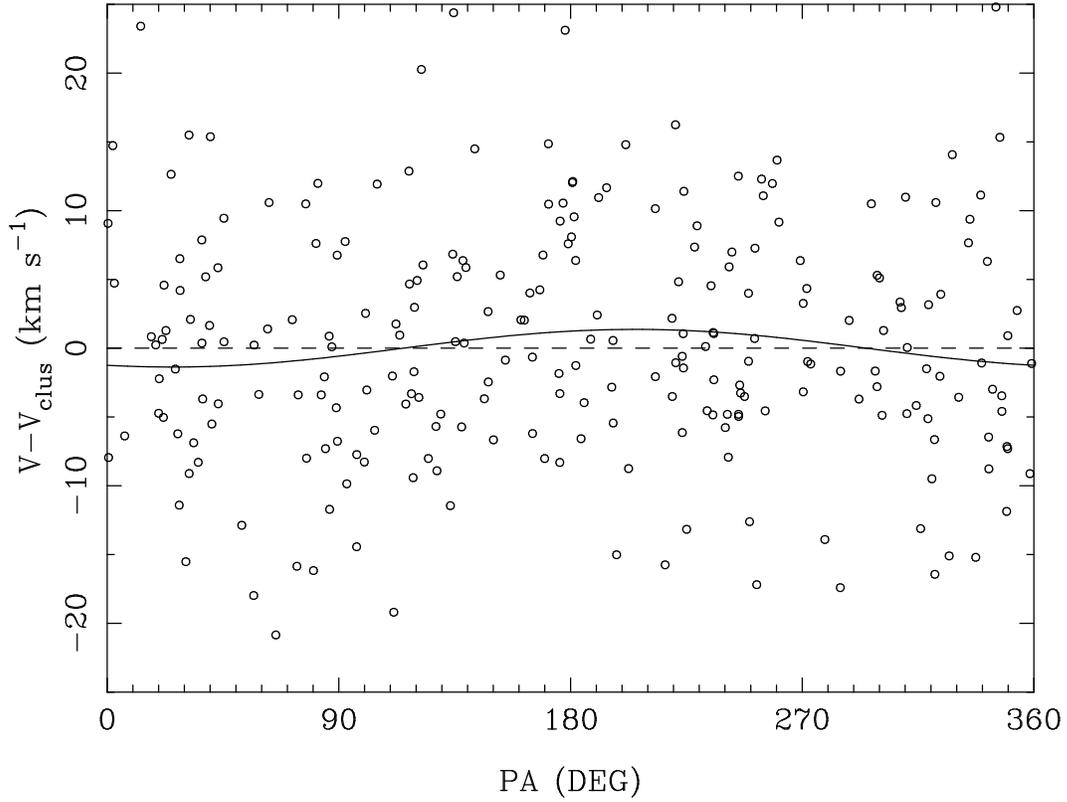

Fig. 6.— Rotation of M15. The circles are the velocities of the stars relative to the cluster mean velocity plotted vs. position angle (from N through E). The solid line is the sine curve that best fits the data.

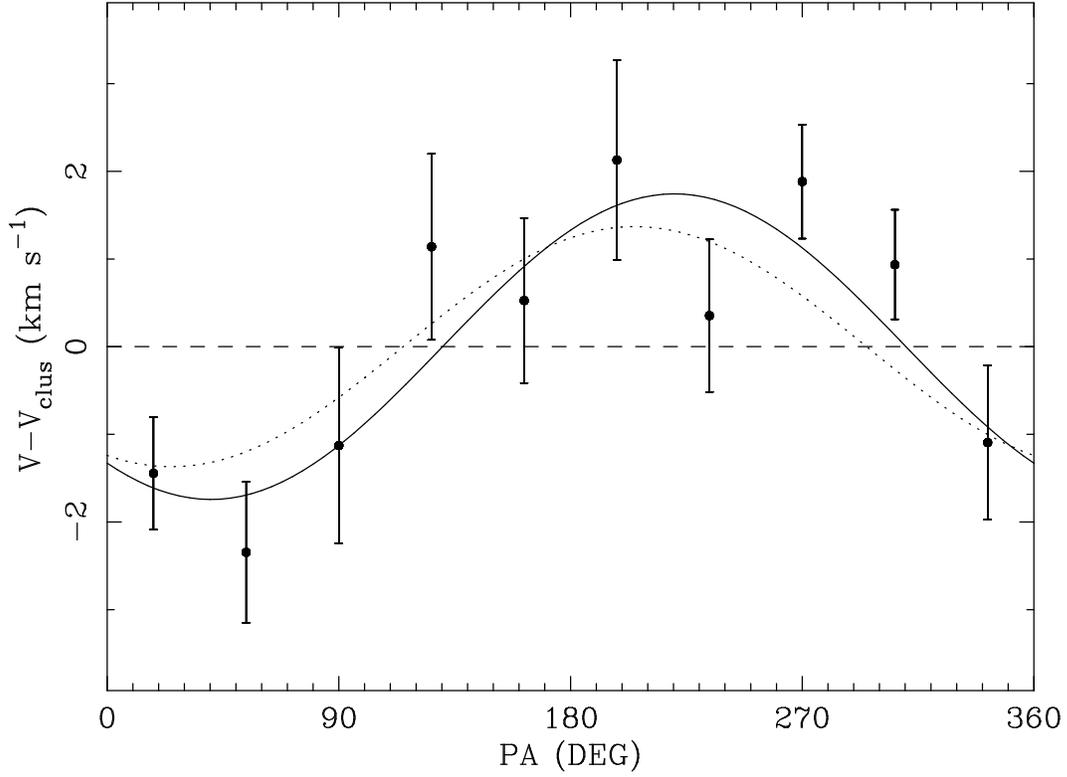

Fig. 7.— Rotation of M15. The open circles are the estimated velocities for the integrated light relative to the cluster mean velocity plotted vs. position angle. The solid curve is the best-fit sine curve and the dotted line is the curve from Fig. 6, which comes from the individual stellar measurements.

## TABLE 1
## M15 STARS

| ID | X($''$) | Y($''$) | $V_{FP}$ | V(km s$^{-1}$) | S | $P_{\chi^2}$ | NOTE |
|---|---|---|---|---|---|---|---|
| AC212 | 2.3 | 0.4 | 13.3 | $-124.2 \pm 1.1$ | | 0.57 | |
| | | | | $-125.9$ | 3.3 | FP92 | |
| | | | | $-123.9$ | 1.2 | PSC | |
| AC224 | $-3.1$ | 0.9 | 14.1 | $-75.4$ | 1.9 | 0.59 | 3 |
| | | | | $-73.9$ | 3.4 | PSC | |
| | | | | $-76.1$ | 2.3 | PSC | |
| AC265 | $-1.8$ | $-2.8$ | 14.2 | $-97.9$ | 3.3 | FP92 | |
| AC254 | $-4.2$ | $-2.3$ | 13.8 | $-102.1$ | 4.2 | FP92 | |
| AC160 | 4.7 | 2.4 | 13.2 | $-97.4$ | 0.8 | 0.64 | |
| | | | | $-99.2$ | 3.3 | FP92 | |
| | | | | $-98.0$ | 3.1 | FP91 | |
| | | | | $-97.8$ | 1.9 | PSC | |
| | | | | $-95.1$ | 1.7 | PSC | |
| | | | | $-98.1$ | 1.2 | PSC | |
| AC253 | $-4.6$ | $-3.5$ | 13.4 | $-107.9$ | 1.6 | 0.62 | |
| | | | | $-108.8$ | 2.4 | FP92 | |
| | | | | $-107.2$ | 2.2 | PSC | |
| AC161 | 6.0 | 1.8 | 13.8 | $-123.9$ | 1.0 | 0.01 | 1 |
| | | | | $-120.9$ | 2.4 | FP92 | |
| | | | | $-132.2$ | 3.0 | FP91 | |
| | | | | $-123.2$ | 1.2 | PSC | |
| AC114 | 5.0 | $-3.9$ | 13.7 | $-116.9$ | 4.1 | FP91 | |
| AC112 | 6.0 | $-3.5$ | 13.6 | $-103.1$ | 2.7 | FP92 | |
| AC3 | $-4.3$ | 5.5 | 13.2 | $-97.4$ | 0.9 | 0.79 | |
| | | | | $-98.2$ | 1.5 | FP92 | |
| | | | | $-96.9$ | 2.7 | FP91 | |
| | | | | $-96.9$ | 1.3 | PSC | |
| AC111 | 7.4 | $-4.1$ | 13.2 | $-117.4$ | 0.8 | 0.69 | |
| | | | | $-119.7$ | 2.8 | FP92 | |
| | | | | $-117.6$ | 2.8 | FP91 | |
| | | | | $-117.2$ | 0.9 | PSC | |
| AC185 | 0.1 | 8.5 | 14.5 | $-98.9$ | 2.4 | FP92 | |
| AC178 | 4.8 | 7.7 | 13.5 | $-92.5$ | 0.9 | 0.29 | |
| | | | | $-95.1$ | 2.2 | FP92 | |
| | | | | $-89.0$ | 3.1 | FP91 | |
| | | | | $-90.5$ | 2.1 | PSC | |
| | | | | $-93.0$ | 1.2 | PSC | |
| K539 | $-6.0$ | $-6.9$ | 13.3 | $-109.1$ | 0.8 | 0.60 | AC6 |
| | | | | $-107.2$ | 3.9 | FP92 | |
| | | | | $-113.1$ | 3.2 | FP91 | |
| | | | | $-108.9$ | 1.2 | PSC | |
| | | | | $-108.9$ | 1.0 | PSC | |
| AC110 | 8.4 | $-4.7$ | 14.3 | $-105.0$ | 1.6 | FP92 | |
| AC773 | 4.9 | 9.1 | 14.4 | $-101.5$ | 2.1 | FP92 | |
| AC291 | $-6.3$ | 8.6 | 15.2 | $-104.1$ | 2.0 | FP92 | |
| AC104 | 10.4 | $-3.9$ | 13.7 | $-97.6$ | 0.8 | 0.83 | 3 |
| | | | | $-98.1$ | 2.4 | PSC | |
| | | | | $-97.6$ | 0.8 | PSC | |
| AC188 | $-2.0$ | 11.1 | 15.3 | $-115.3$ | 3.0 | FP92 | |
| AC247 | $-10.4$ | $-4.8$ | 13.3 | $-95.5$ | 0.7 | 0.37 | |
| | | | | $-94.7$ | 2.0 | FP92 | |
| | | | | $-96.5$ | 1.1 | PSC | |
| | | | | $-94.7$ | 0.8 | PSC | |
| AC100 | 11.9 | 0.6 | 15.2 | $-107.9$ | 4.8 | FP92 | |
| AC616 | 11.0 | $-6.6$ | 13.8 | $-111.6$ | 2.0 | 0.86 | |
| | | | | $-111.1$ | 3.3 | FP92 | |
| | | | | $-111.8$ | 2.5 | FP91 | |
| AC279 | $-7.3$ | 11.2 | 14.4 | $-123.1$ | 1.4 | FP92 | |

NOTE.—
1: Binary star candidates.
2: Not used in dynamics due to radial FP bias.
3: Not used in dynamics due to severe contamination.
4: PSC has 33 measurements for this star.